# Primary damage and mechanical degradation of WTaCrV refractory high-entropy alloy: effects of solid-solution and chemical ordering

*Yihan Wu*[1, *] *Pengfei Yu*[1] *Yaohong Suo*[1] *Lei Zhang*[2, *]

[1] School of Mechanical Engineering and Automation, Fuzhou University, Fuzhou, PR China

[2] Department of Mechanical Engineering, Eindhoven University of Technology, P.O. Box 513, 5600, MB, Eindhoven, the Netherlands

**[*]Corresponding authors:**

wuyihan@fzu.edu.cn (Y Wu), l.zhang9@tue.nl (L Zhang)

**Abstract:** As advanced nuclear reactors demand novel irradiation-tolerant materials, this study investigates the radiation damage and mechanical degradation of the promising WTaCrV refractory high-entropy alloy (RHEA). To isolate complex nanoscale chemical effects, we propose an atomistic modeling strategy comparing Average-Atom (AA), random solid-solution (RSS), and local chemical order (LCO) configurations using newly developed interatomic potentials. Collision cascades simulations reveal that the number of Frenkel pairs follow $N_{\mathrm{RSS}} > N_{\mathrm{LCO}} > N_{\mathrm{AA}}$ at the same radiation dose. While the RSS effect accelerates defect generation due to rugged energy landscapes, LCO enhances lattice cohesion to mitigate radiation damage. Despite more primary defects in the RSS and LCO configurations compared with the AA configurations, the RSS and LCO effects can suppress radiation-induced mechanical degradation. Irradiation severely degrade the homogenized AA model but exert a limited impact on the strength and flow stress of the RSS and LCO models. This exceptional resistance is driven by inherent lattice distortion resulting from interactions among different alloy elements, which outweighs point defect induced lattice disruptions. Moreover, the complex interactions between deformation twins and point defects cause confined plastic flow, elevating flow stress in the RSS and LCO models. The findings provide atomistic guidance for performance assessment of next-generation structural materials for extreme nuclear environments.

## 1. Introduction

As the global demand for sustainable and low-carbon energy intensifies, advanced nuclear reactor systems are being pushed toward unprecedented operational extremes, including higher temperatures and more intense neutron irradiation environments [1, 2]. Traditional nuclear structural materials are rapidly approaching their performance limits under these severe conditions, necessitating the urgent development of next-generation irradiation-tolerant materials. Refractory high-entropy alloys (RHEAs), characterized by multi-principal elements with high melting points, have emerged as promising candidates owing to their inherent lattice distortion, sluggish diffusion, and the synergistic cocktail effect [3, 4]. Among them, the WTaCrV alloy stands out for its exceptional phase stability, remarkable high-temperature strength, and immense potential for irradiation resistance, making it uniquely positioned to tackle challenges in advanced nuclear energy engineering [5-8].

Despite its tempting macroscopic properties, the underlying nanoscopic mechanisms governing the irradiation damage and mechanical performance of the WTaCrV RHEA remain insufficiently understood, hindering its widespread engineering application. Experimental studies have demonstrated the outstanding irradiation tolerance of W-based RHEAs, with negligible irradiation hardening and no observable dislocation loops even after 8 dpa [6, 7]. However, the detailed defect evolution and deformation mechanisms remain largely unclarified. Additionally, it is widely accepted that the fundamental characteristics of RHEAs are heavily dictated by their complex internal chemical environments, specifically the random solid-solution (RSS) and local chemical order (LCO) effects. The RSS effect originates from the inherent chemical disorder of multi-principal elements randomly occupying lattice sites, which induces significant lattice distortion and localized strain fields at the atomic scale [9, 10]. The LCO effect, by contrast, arises from enthalpic interactions among constituent elements, leading to preferential elemental pairing [11-13]. As pointed out by preliminary studies [2, 12, 14, 15], the RSS and LCO effects critically influence defect formation, migration, and plastic flow during irradiation and loading. However, a major scientific bottleneck in current research is that both RSS and LCO effects manifest at the nanometer or even sub-nanometer scales [11, 13]. This spatial restriction makes it challenging to experimentally measure, isolate, and independently study their individual impacts on material behaviors. This knowledge gap directly impedes the rational design and performance optimization of W-based RHEA for nuclear applications.

To overcome the limitations of experimental characterization, atomistic simulations such as molecular dynamics (MD) are invaluable for tracking primary collision cascades, nanoscale plasticity, and the associated RSS and LCO effects. Nevertheless, computational exploration of the quaternary W-Ta-Cr-V system is currently severely hindered by the lack of suitable modeling strategies and robust interatomic potentials capable of explicitly decoupling these two effects. In conventional full-element MD models, the RSS and LCO effects inherently coexist and are convoluted within a single configuration, making it difficult to attribute observed behaviors to either factor alone. Although some studies attempt to isolate RSS and LCO effects by comparing the properties of HEAs with those of pure constituent metals [2, 16], such comparisons are challenged by the inherent differences in the basic properties and resulting behaviors of alloys and metals. To address this challenge, the average-atom (AA) approach has been proposed, in which an 'artificial' single-element material with compositionally averaged properties replaces the multi-principal alloy [17]. This approach has recently been applied to study radiation damage and indentation responses of FeNiCrCo-based HEAs [2, 18] and to investigate the deformation mechanisms of HfNbTaZr-based RHEAs [4, 14]. By comparing the distinct responses of AA, RSS, and LCO configurations, the independent roles of chemical randomness and short-range ordering can be isolated. However, the reliability of the AA model largely depends on the accuracy of the full-element interatomic potentials, as it is constructed through a weighted combination of the constituent metal potentials. Therefore, developing a high-precision interatomic potential that accurately captures the energetic, structural, mechanical, and defect properties of the W-Ta-Cr-V quaternary system is essential to enable such a modeling strategy.

To fill these research gaps, the atomistic modeling strategy is applied to the WTaCrV RHEA

to systematically isolate and quantify the individual and coupled effects of RSS and LCO by comparing AA, RSS, and LCO configurations. Interatomic potentials for the WTaCrV RHEA are developed based on density functional theory (DFT) reference data. By systematically simulating primary collision cascades and subsequent uniaxial tension, we elucidate how nanoscale chemical environments influence defect evolution and mechanical degradation. The remainder of this paper is organized as follows: Section 2 outlines the computational methods, including the potential development and simulation settings; Section 3 presents and discusses the main findings of this work, including the basic properties and local chemical ordering of WTaCrV, radiation damage evolution, and mechanical degradation behaviors.

## 2. Computational details

### 2.1 Interatomic potential development

As pointed out in previous studies, RSS and LCO effects can cause nanostructural distortion and strain energy accumulation in the RHEA, which further modulate the energy barriers of materials behaviors such as diffusion, plastic flow or defects nucleation [4, 10, 12, 19]. However, both RSS and LCO take effect within nanometer scale [11, 13], thereby challenging to be experimentally measured. To overcome this barrier, a novel atomistic modelling strategy is proposed, as schematically displayed in Fig. 1a. The RSS effect is a result from the complex mutual interactions among various constituent metals in the RHEA. Therefore, the RSS effects can be isolated if we design an elemental material containing only one 'artificial' element (i.e., the A-atom or AA [15, 17, 18]), whose properties are identical to the macroscale averaged values of the RHEA. Besides, the LCO effect can be incorporated using Monte Carlo (MC) optimization [8], which rearrange distribution of elements according to their mutual affinity. By comparing materials responses of AA, RSS and LCO configurations, the roles of atomic scale chemical environments on radiation damage and mechanical degradation of WTaCrV RHEA can be revealed.

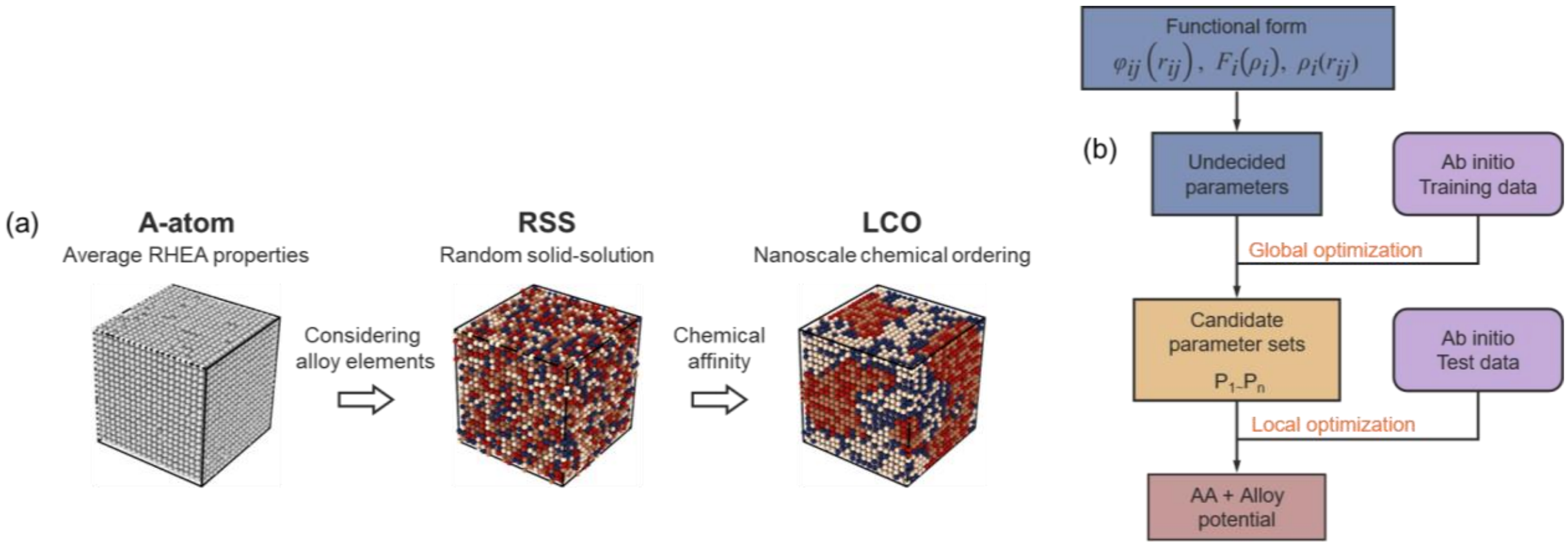

**Fig. 1** (a) Modelling strategy to separate RSS and LCO effects. (b) Flowchart of interatomic potential development.

The modelling strategy in Fig. 1a requires suitable interatomic potentials, which must be first developed. In this study, the workflow of potential development is detailed in Fig. 1b. We select the Embedded Atom Method (EAM) that has been validated to balance accuracy and computational efficiency in simulating metallic materials. The general functional form of EAM can be expressed as [20]

$$E = \frac{1}{2}\sum_{i=1}^{N}\sum_{j=i_1}^{i_M}\varphi_{ij}(r_{ij}) + \sum_{i=1}^{N}F_i(\rho_i), \tag{1}$$

where $\varphi(r)$, $F(\rho)$ and $\rho(r)$ are the pair interaction, the embedded function and the electron density. In this study, the widely utilized Zhou and Wadley functional form of EAM [21] is adopted, given as:

$$\varphi(r) = \frac{A\exp\left[-\alpha\left(\frac{r}{r_e}-1\right)\right]}{1+\left(\frac{r}{r_e}-\kappa\right)^m H\left(\frac{r}{r_e}-\kappa\right)} - \frac{B\exp\left[-\beta\left(\frac{r}{r_e}-1\right)\right]}{1+\left(\frac{r}{r_e}-\lambda\right)^n H\left(\frac{r}{r_e}-\lambda\right)}, \tag{2}$$

$$F(\rho_i)=\begin{cases}\sum_{i=0}^{3}F_{ni}\left(\frac{\rho_i}{\rho_n}-1\right)^i, & \rho<\rho_n,\ \rho_n=0.85\rho_e\\ \sum_{i=0}^{3}F_i^-\left(\frac{\rho_i}{\rho_e}-1\right)^i, & \rho_n\leq\rho<\rho_e\\ \sum_{i=0}^{3}F_i^+\left(\frac{\rho_i}{\rho_e}-1\right)^i, & \rho_e\leq\rho<\rho_o,\ \rho_n=1.15\rho_e\\ F_e\left[1-ln\left(\frac{\rho}{\rho_s}\right)^\eta\right]\left(\frac{\rho}{\rho_s}\right)^\eta, & \rho_o\leq\rho\end{cases}, \tag{3}$$

$$\rho_i=\sum_{i=1}^{N}f_i(r_{ij}),\ f_i(r_{ij})=\frac{f_e\exp\left[-\beta\left(\frac{r}{r_e}-1\right)\right]}{1+\left(\frac{r}{r_e}-\lambda\right)^n}. \tag{4}$$

Eqs. 2~4 give 23 adjustable parameters for each element. Considering a total of four metal elements (W, Ta, Cr and V) and their mutual interactions, 119 internal parameters need to be determined for a reliable alloy potential that accurately predicts reference properties of pure metals, intermetallic compounds and alloys. Here, we use density functional theory (DFT) to calculate these reference data, including energetic, structural, mechanical and defect properties (the comparison of MD and DFT data are detailed in section S1 of the supplementary material). The DFT calculations are performed using the Quantum Espresso package [22]. The pseudopotentials are constructed with the ultrasoft/PAW method and the exchange correlation is described by the PBE-GGA form [23, 24]. The Brillouin Zone is sampled using Monkhorst-Pack grids; the *k* spacing is set to < 0.03 $Å^{-1}$.

Subsequently, all internal parameters are adjusted using a homemade package performing distributed breeder genetic algorithm (DBGA) [25, 26], which searches the entire variable space and efficiently approaches to the global optimal. Eighty percent of DFT data are included in this training process. Then, 5% of the parameter sets with lowest loss function values are chosen; a local SIMPLEX optimization is conducted to further minimize the prediction errors towards the remaining 20% of DFT data. Finally, the optimal parameters for the full element system are obtained, hereafter referred to as the Alloy potential (all parameters are listed in section S1 of the supplementary material).

The A-atom (AA) potential can be directly constructed based on the developed W, Ta, Cr and V potentials as [17]:

$$\langle E_0 \rangle = \sum_i F^{AA}(\bar{\rho}_i) + \frac{1}{2}\sum_{i,j\neq i} V_{ij}^{AA}, \tag{5}$$

with $$F^{AA}(\bar{\rho}_i) = \sum_X c_X F^X(\bar{\rho}_i),\ V_{ij}^{AA} = \sum_{X,Y} c_X c_Y V_{ij}^{XY},\ \bar{\rho}_i = \sum_{j\neq i}\sum_X c_X \rho_{ij}^X. \tag{6}$$

In Eqs. 5~6, $c_X$ denotes nominal concentration of an element $X$ in the RHEA (namely, 25 % in equimolar WTaCrV). This combination strategy is proven reliable in previous studies on FeCoNiCr-based FCC structured alloys [2, 18].

To improve the accuracy of Alloy and AA potentials in simulating collision cascades, the pair interactions within 0.5 Å are modified by the Ziegler-Biersack-Littmark (ZBL) potential [27]:

$$V_{ij}^{ZBL} = \frac{1}{4\pi\varepsilon_0}\frac{Z_i Z_j e^2}{r_{ij}}\Phi(r_{ij}/a), \tag{7}$$

where $e$ is the charge of an electron, $\varepsilon_0$ is the permittivity in vacuum, $Z$ is the atomic number, $\Phi$ is the screening function. A connection function [5, 25, 28] is used within the range of 0.5~1.0 Å to ensure smooth transition between EAM and ZBL.

**2.2 Simulation settings**

To conduct MD simulations on primary irradiation damage in the WTaCrV RHEA, BCC-structured single crystalline AA and RSS configurations are first created with dimensions of $60a\times60a\times60a$ where $a$ is the equilibrium lattice constant (the choice of model size is detailed in section S2.1 of the supplementary material). Periodic boundary conditions are applied in all directions. The two types of configurations (namely, AA and RSS configurations) are thermally equilibrated at 300 K for 500 ps, using an NPT ensemble to release internal stress. Thereafter, a hybrid MC/MD optimization is performed on the RSS configurations, during which atomic species are switched to reduce the potential energy of the system [8]. The swap trials are accepted according to Metropolis criterion; a simulated annealing strategy [4] is utilized to gradually reduce temperature from 1200 K to 300 K. The MC optimized models are equilibrated for another 500 ps in NPT ensemble; the final models are denoted as LCO configurations.

During cascade simulations, boundary layers with a thickness of 10.0 Å is endowed the NVT ensemble along with the Nosé-Hoover thermostat to absorb the collision energy and cool the system back to 300 K [19, 29], while the central simulation domain are under the NVE ensemble. A tungsten atom is randomly selected as the primary knock-on atom (PKA) and is then moved to the center of simulation domain by shifting all atoms across periodic boundaries. The PKA is then given an initial kinetic energy ranging from 5 keV ~ 40 keV with the velocity along the <135> direction to avoid channel effects [29, 30]. The system is subsequently relaxed for 200~500 ps, as listed in Table 1. For each configuration under a specific PKA energy, 15 independent simulations are conducted to represent the statistical fluctuation. An adaptive timestep (from $10^{-4}$ fs to 1.0 fs) is utilized; the timestep is adjusted so that each atom did not move more than 0.05 Å per step.

**Table 1** The PKA energies considered in this study and the associated relaxation time after initial collision.

| **PKA energy** | **Relaxation time (ps)** |
|---|---|
| 5 keV | 200 ps |
| 10 keV | 300 ps |
| 20 keV | 400 ps |
| 30 keV | 400 ps |
| 40 keV | 500 ps |

To investigate the RSS and LCO effects on the irradiation induced mechanical degradation, the configurations after PKA collisions are replicated two times along the *z* direction. The *x* and *y* directions are switched to shrink-wrapped boundary conditions to create a nanopillar with free surfaces. The nanopillars are then uniaxially stretched via displacement loading at 300 K. The stress evolution is recorded to analyze mechanical responses.

In this study, simulations are performed using the LAMMPS simulation package [31]; MD results are visualized using the OVITO software [32]. Crystalline structures are analyzed using the Polyhedral template matching (PTM) algorithm [33]; point defects are identified using the Wigner-Seitz analysis tool [2] and dislocations are analyzed via DXA algorithm [34]. In all simulations except collision cascades, the timestep is fixed as 1.0 fs; the loading strain rate is set as $10^8$/s which is typical in MD studies [35] (an analysis on strain rate sensitivity is given in section S2.2 of the supplementary material).

## 3. Results

### 3.1 Basic properties and local chemical order (LCO)

In this section, some basic properties of the WTaCrV RHEA are computed using the developed interatomic potentials and compared with DFT data, in order to demonstrate their reliability. Besides, the nanoscale chemical ordering is revealed by hybrid MC/MD simulations, which is the prerequisites for investigating LCO effect. Furthermore, the RSS- and LCO-influenced formation energies of point defects are also analyzed for subsequent radiation damage study.

Fig. 2a displays the cohesive energy, lattice parameter and elastic constants calculated using DFT, Alloy and AA potentials. Results suggest that both developed potentials reasonably predict these properties of WTaCrV RHEA, consistent with DFT data. Besides, these properties of WTaCrV are found to obey the rule of mixing (ROM), i.e., the alloy properties can be described using those of constituent metals:

$$Q_{\mathrm{RHEA}} = \sum_{i=1}^{i=N} c_i Q_i, \tag{8}$$

where $Q$ denotes a specific property, $c$ is the nominal concentration and $i$ = 1~4 represents W, Ta, Cr and V. The consistence with ROM is analogous to a broad range of other HEAs with either BCC or FCC structures [4, 36, 37], indicating it may be a universal phenomenon that validates the concept of the AA method (Eq. 6). Nonetheless, consistence of static properties with ROM does not necessarily represent that dynamic material behaviors are also manifested as averages of their constituents. The RSS and LCO effects are non-negligible, as will be further discussed in the following sections.

In addition to the basic properties in Fig. 2a, stacking fault energy is another important property as it is closely related to nanoscale plastic deformation mechanisms (i.e., dislocation slip, deformation twinning, etc.) [4, 38, 39]. Fig. 2b shows the generalized stacking fault energy (GSFE) curves for two typical slip systems of BCC crystal (namely, <110>(110) and <111>(112)). GSFE predicted using AA and Alloy potentials exhibits consistent values, which is also in agreement with DFT calculated GSFE curves. The correctly predicted elastic constants and GSFE curves (Figs. 2a and 2b) suggest the applicability of our potential in simulating mechanical behaviors of WTaCrV RHEA under various loading conditions [4, 12, 39]. Further performance tests of the developed potentials (e.g., mechanical properties of pure metals and equation of states for binary and ternary compounds) are detailed in section S1 of the supplementary material.

More detailed performance evaluations (e.g., the mechanical properties of pure metals and the equations of state for binary and ternary compounds) of the developed potentials are presented in Section S1 of the Supplementary Material.

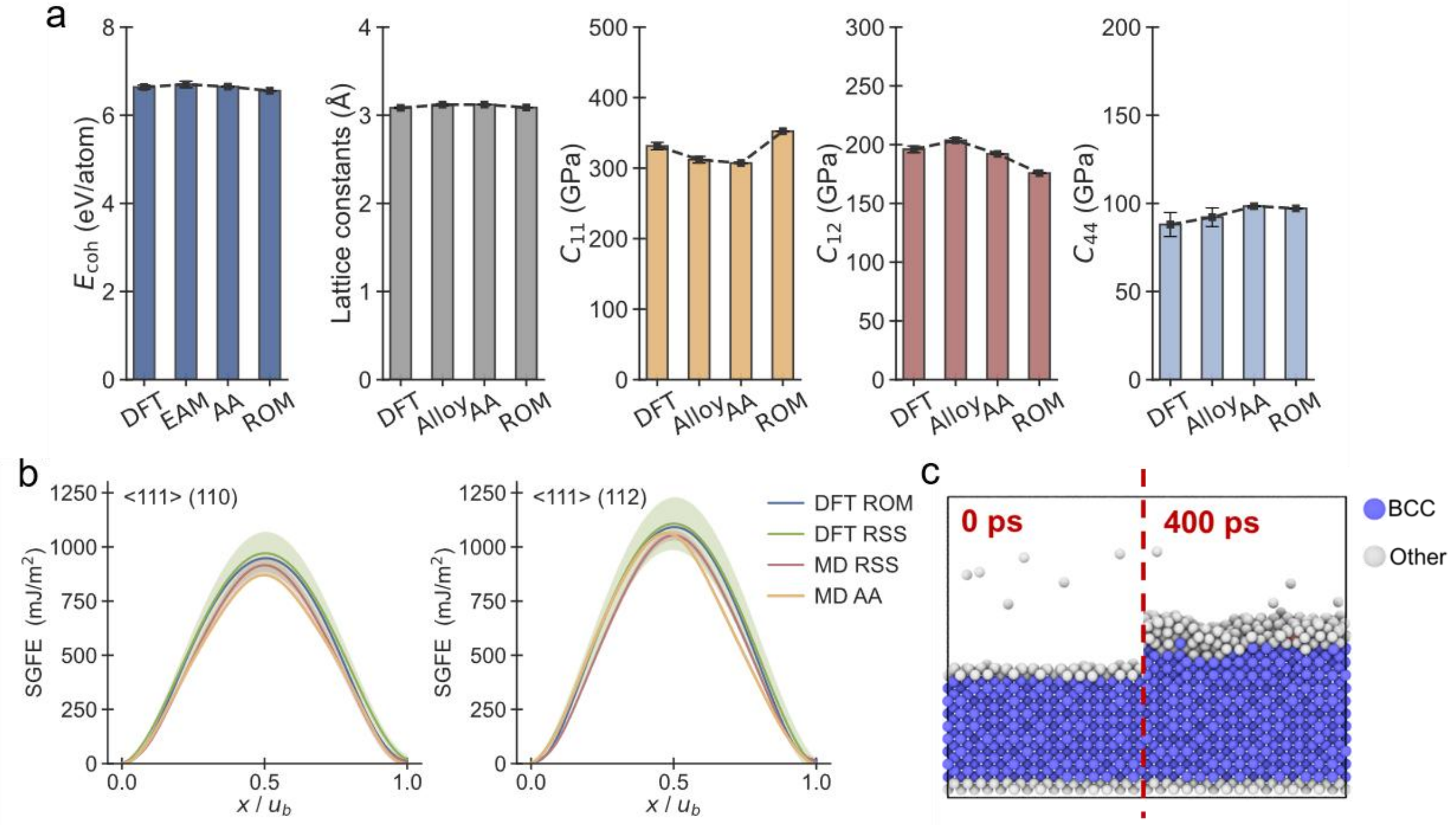

**Fig. 2** (a) Cohesive energy $E_{\mathrm{coh}}$, lattice constant $a$, elastic constants $C_{11}$, $C_{12}$, $C_{44}$ of the WTaVCr RHEA, calculated via DFT, Alloy potential, AA potential and rule of mixing. (b) Comparison of GSFE curves calculated using different methods. Shaded bands represent the variation range. (c) Atomic deposition process predicted using the developed potential.

In addition to the static properties, it is also necessary to test the potentials' ability to retain structural stability in dynamic simulations under finite temperatures. Previous studies have indicated that, with highly flexible function forms and massive tunable parameters, complex force fields (e.g., machine learning force fields and some bond-order force fields) may fail by structural collapse in dynamic simulations although they reproduce static properties with DFT accuracy [4, 40]. Thus, we conduct atomic deposition simulation under room temperature as shown in Fig. 2c. Both Alloy and AA potentials are able to describe the process where vaporized atoms deposit on substrate and spontaneously grow into the ground state BCC structure. This indicates that the developed potentials are applicable in simulating RHEA in extreme conditions far from the equilibrium state, e.g., radiation cascade and shock compression. Results in Fig. 2 validate the accuracy and wide applicability of the EAM functional form, not to mention its inherent high computational efficiency [41].

Subsequently, we study the nanoscale chemical ordering using the developed Alloy potential through hybrid MC/MD optimization. The optimized RHEA configuration is displayed in Fig. 3a, in which apparent LCO can be observed. The ordering patterns can be captured and quantified using the Warren-Cowley (WC) order parameters, given as [12]

$$\alpha_{ij}^{n} = \left(p_{ij}^{n} - C_j\right) / \left(\delta_{ij} - C_j\right), \tag{9}$$

where $n$ represents the $n$-th nearest-neighbor shell surrounding a central atom $i$, $p_{ij}$ is the average probability of finding a $j$ atom around the $i$ atom within the $n$-th shell, $C_j$ is the macroscale concentration and $\delta_{ij}$ is the Kronecker delta function. For pairs of different elements ($i \neq j$), negative values denote clustering. The calculated WC parameters $\alpha^{1}{}_{ij}$ are shown in Fig. 3b. The negative values for W-V and Ta-Cr pairs suggest strong attraction, which may be attributed to the large negative mixing enthalpies among these two pairs of elements as reported in previous studies [8, 42]. However, this LCO pattern within single crystalline RHEA (Fig. 3a) seems inconsistent with some experimental observations, which report phase separation into W-Ta-enriched and Cr-V-enriched regions [6, 7]. A recent machine learning study has pointed out that such discrepancy may result from the significant difference in lattice constants between WTa and CrV binary compounds [8].

Therefore, huge misfit strain will generate if the W-Ta-enriched and Cr-V-enriched regions share a coherent interface. We thereby construct a model as shown in Fig. 3c, where a spherical BCC precipitate (*a*=2.7Å) is inserted into RHEA BCC lattice (*a*=3.15Å), creating a semi-coherent interface to prevent accumulation of strain energy. As revealed in Fig. 3c, after MC optimization, the W-Ta and Cr-V clustering is successfully reproduced in agreement with experiments. Results in Fig. 3 demonstrate the developed potential can correctly predict interaction among alloy elements and thus the LCO phenomenon. It can be utilized to investigate the RSS and LCO effects on radiation damage and mechanical behavior.

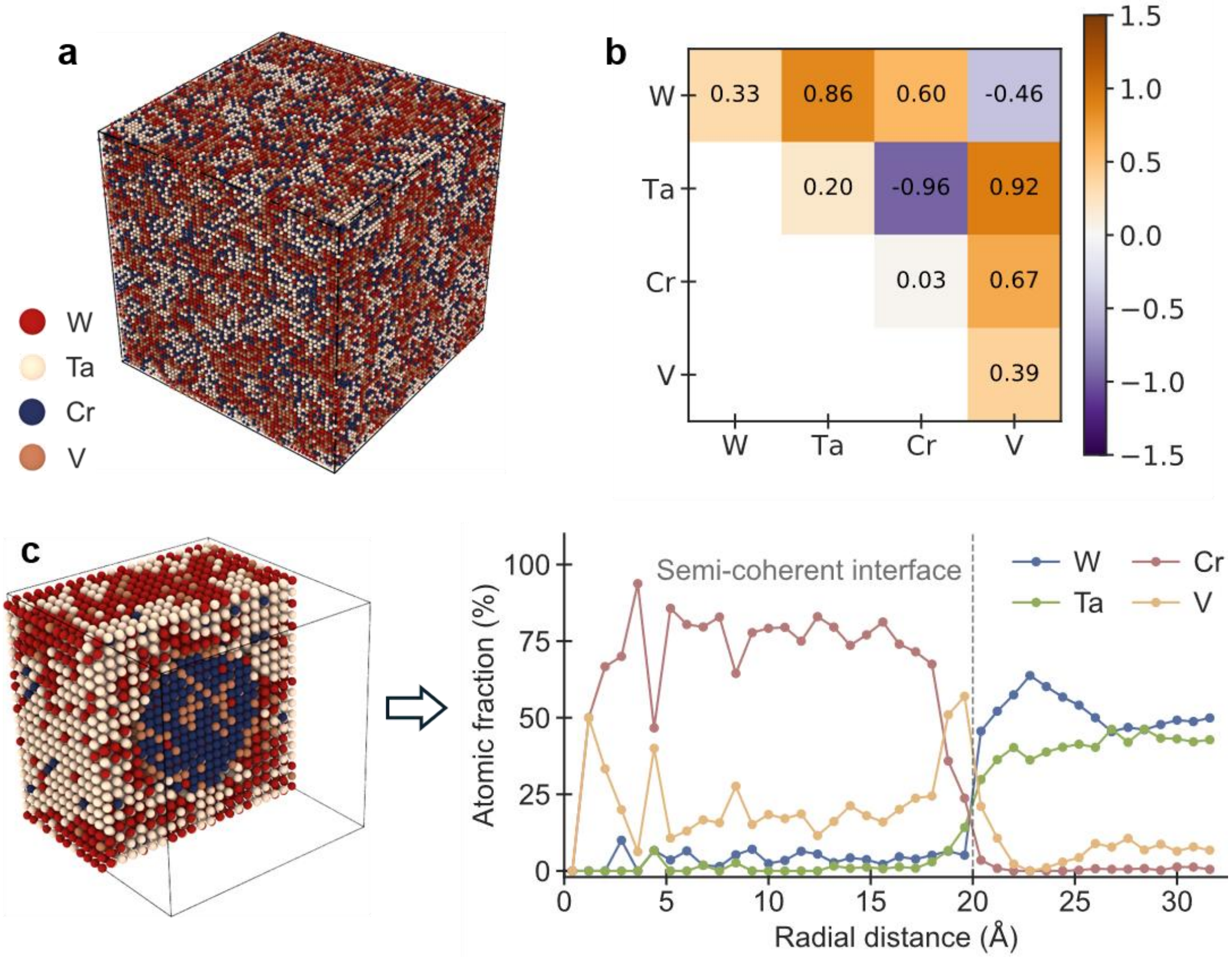

**Fig. 3** (a) The MC-optimized single crystalline RHEA with internal LCO. (b) The corresponding WC parameters of different element pairs. (c) Atomic configuration of $W_{38}Ta_{36}Cr_{15}V_{11}$ with a BCC precipitate ($R$ = 20 Å) after MC optimization. The alloy composition is selected based on a previous experiment [7]. A semi-coherent interface is created by inserting a spherical precipitate with a smaller lattice constant. The related compositional profile is also shown.

Finally, the RSS and LCO effects on the formation energy of point defects (i.e., vacancies and interstitials) are analyzed. As shown in Figs. 4a~4b, both vacancy and interstitial formation energies of the RSS RHEA exhibit broad distributions, which results from the complex internal interactions among elements induced by the high-entropy effect [43]. The average formation energies of vacancies and interstitials are $E_v^f$ = 3.042 eV/atom and $E_i^f$ = 5.028 eV/atom, respectively. The $E_i^f$ is calculated via energy minimization of randomly introduced single interstitials. The defect formation energies are close to previously reported DFT values [44]. Besides, the average $E_v^f$ and $E_i^f$ of RSS models are consistent with AA predictions. However, the consistent average defect formation energies do not necessarily represent identical resistance against radiation damage; the effect of interactions among different elements will be detailed in the following sections. Moreover, Figs. 4c and 4d show that the average $E_v^f$ and $E_i^f$ of LCO configurations increase to 3.501 eV/atom and 8.538 eV/atom, respectively. The increased defect formation energies indicate that LCO may suppress nucleation of point defects during cascades. Notably, the LCO-induced increase in $E_i^f$ is larger than that in $E_v^f$, which can be understood by decomposing the defect formation energy within the EAM framework: $\Delta E_f = \Delta E_{pair} + \Delta E_{embed}$, where the pair-interaction term (corresponding to the pairwise interaction $\varphi_{AB}$) dominates [20]. For creating a vacancy, the energy cost arises primarily from removing an atom at the equilibrium nearest-neighbor distance ($d_{vac}^{equi} \approx 2.7$ Å), where all ten pair potentials nearly coincide (Fig. 4e, inset). For an interstitial, the inserted atom forces its neighbors into the short-range repulsive part of $\varphi_{AB}$ (e.g., the octahedral site lies only $d_{int}^{oct} \approx 1.56$ Å from the nearest neighbors). In this steeply repulsive regime, the differences among all 10 $\varphi_{AB}$ functions are much more significant at $d_{int}^{oct}$, indicating the interstitial formation more sensitive to the local chemical environment than that of the vacancy.

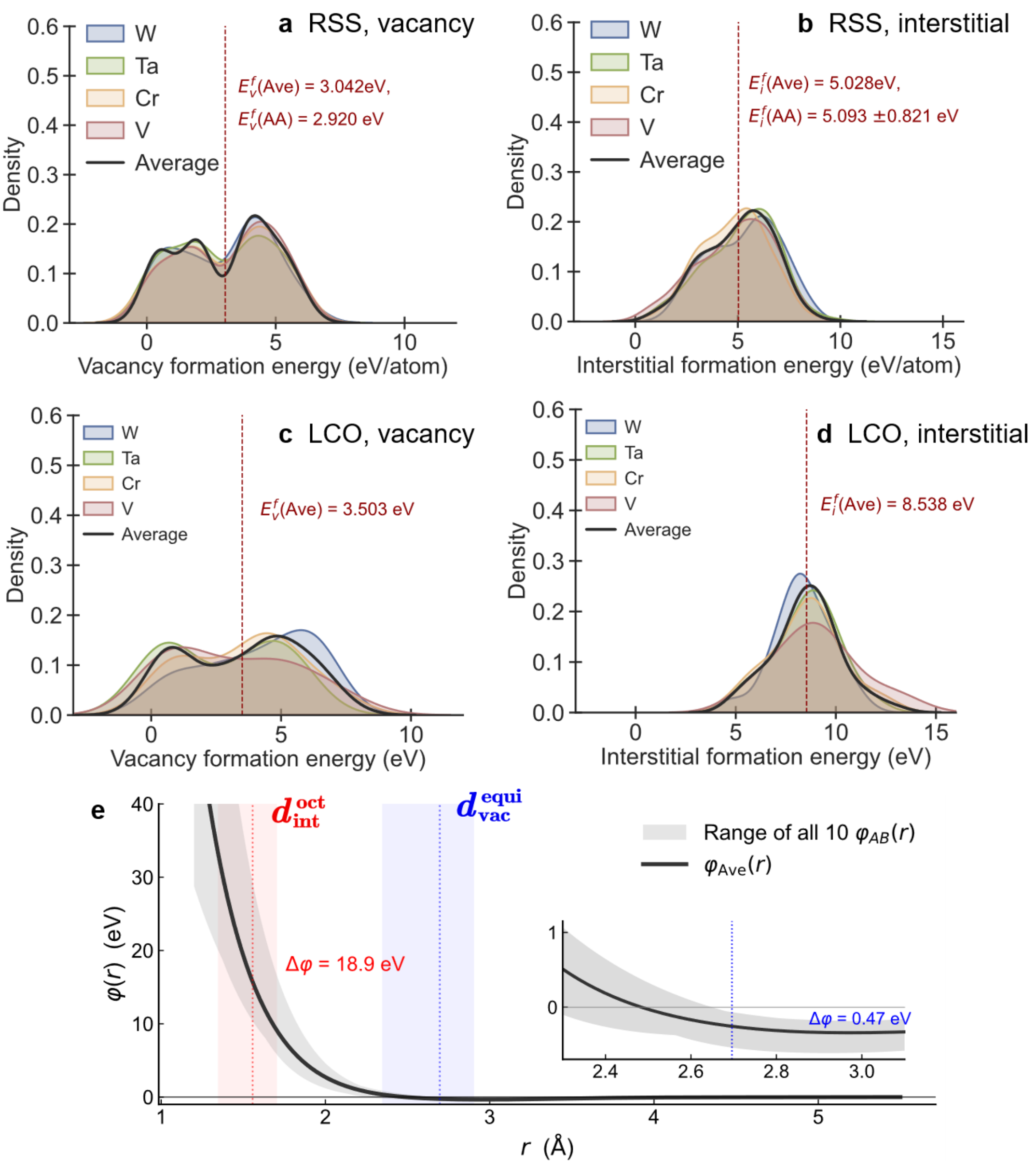

**Fig. 4** (a)~(d) Kernel density estimation (KDE) [45] on point defect formation energies of AA, RSS and LCO configurations of the WTaCrV RHEA. Each curve is calculated from 500 independent samples. Dashed lines represent the mean values of the probability distribution. (e) Pair-potential envelope for WTaCrV. The black curve represents the average of all 10 pairwise functions $\varphi_{AB}$, and the shaded area indicates the range. The inset shows a magnified area near the equilibrium position (~2.7 Å).

### 3.2 Radiation damage

In this section, the RSS and LCO effects on primary damage of WTaCrV RHEA are systematically studied. The time evolution of Frenkel pairs (FPs) in AA, RSS and LCO configurations under collision cascades with varying PKA energy is shown in Fig. 5. For all collision cases, the number of FPs promptly rises at first and gradually reduces after the thermal spike. Among all PKA energies, the number of FPs in the three configurations consistently follows the sequence: $N_{\mathrm{RSS}} > N_{\mathrm{LCO}} > N_{\mathrm{AA}}$. It can thus be inferred that the RSS effect weakens the damage resistance, while LCO effect enhance the resistance. In fact, the reduced damage resistance in RSS RHEA can be explained by the defect formation energy in Fig. 4. Unlike the AA models, the complex interactions among metal elements in the RSS configurations result in rugged energy landscape for formation, migration, agglomeration and annihilation of point defects [9, 11, 12, 43], as evidenced by the broad energy distribution in Figs. 4a and 4b. Consequently, there exist nanoscale 'weak' regions with lower defect formation energies in RSS alloy, serving as preferable nucleation sites for vacancies and interstitials. This explains the higher number of FPs in RSS models compared with AA models under an identical PKA level. On the other hand, the existence of LCO tend to enhance lattice cohesion [46, 47] and stabilize the BCC crystalline structure (the average cohesive energy is -6.92 eV/atom for LCO RHEA compared to -6.70 eV/atom in RSS RHEA). Therefore, both the resistance against FP generation and the ability for recombination of FPs are improved, leading to lower number of defects during primary damage stage compared with the RSS models.

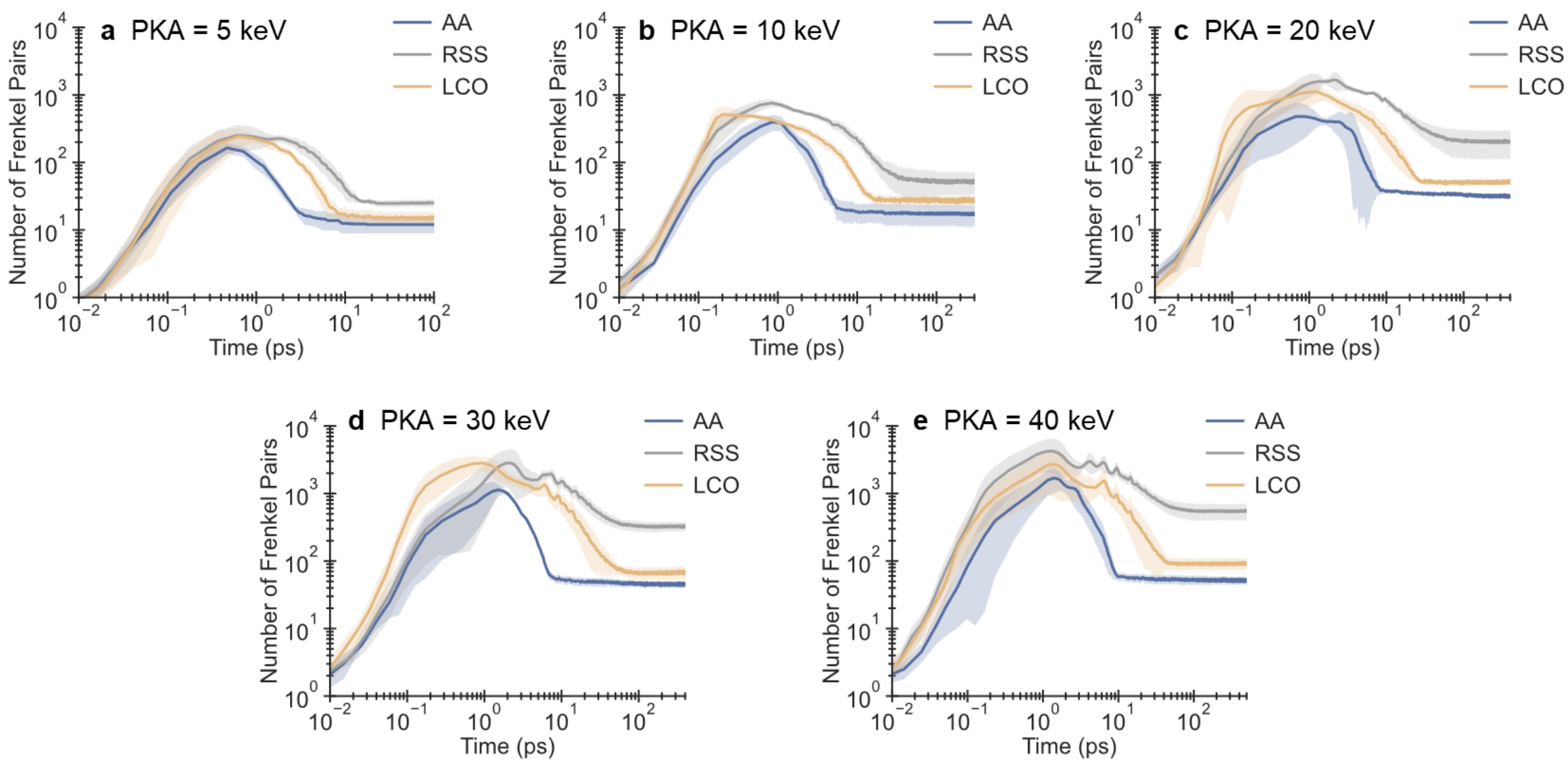

**Fig. 5** The evolutions of average numbers of Frenkel pairs in AA, RSS and LCO configurations as functions of time for PKA energies of (a) 5 keV, (b) 10 keV, (c) 20 keV, (d) 30 keV, and (h) 40

keV. The solid lines represent the average values over 15 independent samples, and the shaded band indicates the range.

The peak number and the surviving number of FPs as functions of PKA energy for AA, RSS and LCO configurations are shown in Figs. 6a and 6b. For three configurations, the numbers of FPs increase with increasing PKA energy. The RSS models exhibit highest damage level, followed by the LCO models. The AA models show lowest numbers of FPs at thermal spike and after relaxation. In addition, the difference among FP numbers becomes more significant as PKA energy increases. Under 40 keV PKA energy, the peak number of FPs in RSS models is nearly twice higher than that in the AA models, while the surviving FP number in RSS models is almost ten times as many as that in the AA models. Besides, based on data in Figs. 6a and 6b, the recombination rate $\eta$ of FPs can be calculated. Fig. 6c shows that for all RHEA configurations, $\eta$ ranges from 89 % to 97 %, indicating good radiation tolerance of the WTaCrV RHEA. Under all studied PKA energy, RSS configurations exhibit lowest recombination rate. Meanwhile, the recombination rates of FPs are similar for AA and LCO models, regardless of PKA energy. Results in Fig. 6 also illustrate the opposite effects of RSS and LCO, i.e., RSS weakens the damage resistance and radiation tolerance while LCO enhances these properties.

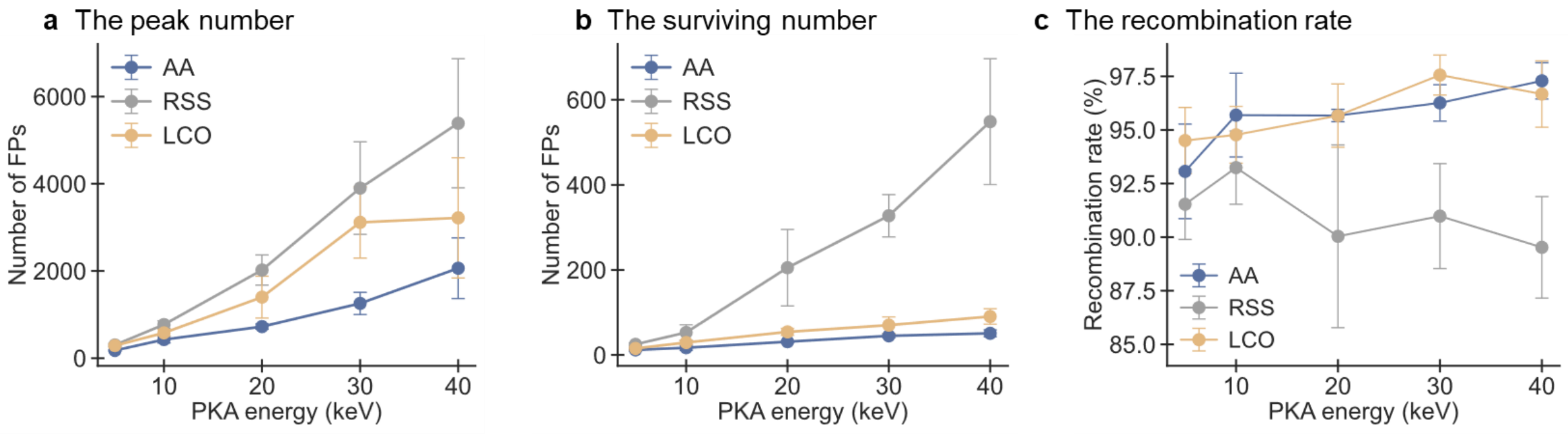

**Fig. 6** (a) The number of FPs at the thermal spike and (b) the number of surviving FPs as functions of PKA energy. The recombination rate is calculated using $\eta = (N_{\mathrm{peak}} - N_{\mathrm{final}}) / N_{\mathrm{peak}}$ and displayed in (c).

Subsequently, we investigated how the surviving defects after PKA collision distributed in RHEA with different local chemical environments. The size distributions of interstitial clusters are shown in Fig. 7. Notably, the distributions of vacancy clusters and interstitial clusters are similar and therefore only the latter is presented. Under 5 keV PKA, interstitial clusters in AA, RSS and LCO models exhibit similar numbers and contain only two point defects. As PKA energy increasing, both the number and size of clusters promptly increase in RSS models. In contrast, the total numbers of interstitial clusters in the AA and LCO models mildly increase. Even subjected to 40 keV PKA collision, the largest cluster size remains less than 10 interstitials in these two models. Furthermore, it is worth noting that no dislocation loop is observed in all cases, suggesting excellent resistance of WTaCrV against defect accumulation and mechanical degradation (which will be discussed in the later sections).

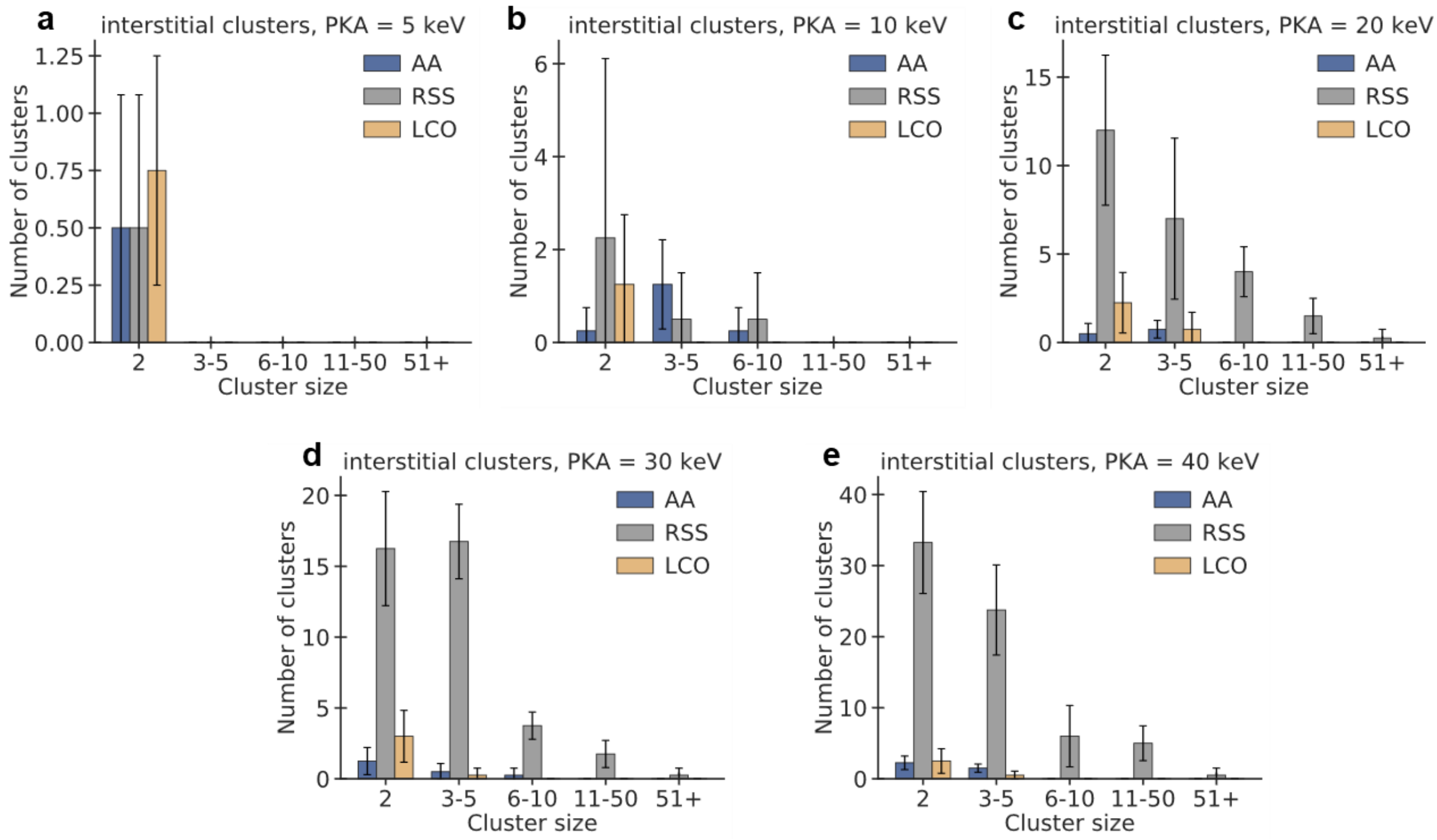

**Fig. 7** Distributions of average numbers of interstitials clusters for AA, RSS and LCO configurations, after (a) 5 and (b) 10 keV, (c) 20 keV, (d) 30 keV and (e) 40 keV PKA cascade.

The time evolution of temperature in NVE regions are shown in Fig. 8. Clearly, the temperature decreases most quickly in the AA configurations under PKA collisions. Besides, as PKA energy increases, the difference in cooling rates between RSS and LCO models becomes more apparent. It takes longest time for RSS models to recover the steady state, which is consistent with the defect evolution behavior in Fig. 5. Results in Fig. 8 suggest that RSS effect weakens the ability to dissipate the collision energy, while LCO effect slightly enhances the energy dissipation capability.

Additionally, the internal temperature patterns at thermal spike and during relaxation are detailed in Fig. 9. It is revealed that the cascade-induced thermal spike is restricted within a relatively smaller region in the RSS model, while the incident energy spreads more widely in AA and LCO models. Besides, the overall temperature at 65 ps is relatively lower in the AA model compared to the RSS and LCO models, evidencing its highest cooling rate. Moreover, some studies on high-entropy alloys or ceramics have reported that the high configurational entropy in multi-principal materials can cause reduced electron mean free path and a decrease in electrical and thermal conductivity [25, 28, 30]. This further leads to slower energy dissipation after collision. Such a longer lifetime of thermal spikes can favor recombination of local FPs at the core of the cascade, thus suppressing the formation of irradiation defect clusters [2, 29, 48]. However, larger numbers and sizes of clusters in RSS models (Fig. 7) indicate that, for the WTaCrV RHEA, a slower cooling rate resulting from the RSS effect does not necessarily facilitate damage recovery. Also, it is usually attributed to the 'high-entropy' effect that causes the slower energy dissipation [49], but in real RHEAs consisting of multi-principal components, the 'high-entropy' effect is always the combined consequence of RSS and LCO effects. Fig. 8 demonstrates that the RSS effect (i.e., chemical disorder) is the main cause of such slow dissipation.

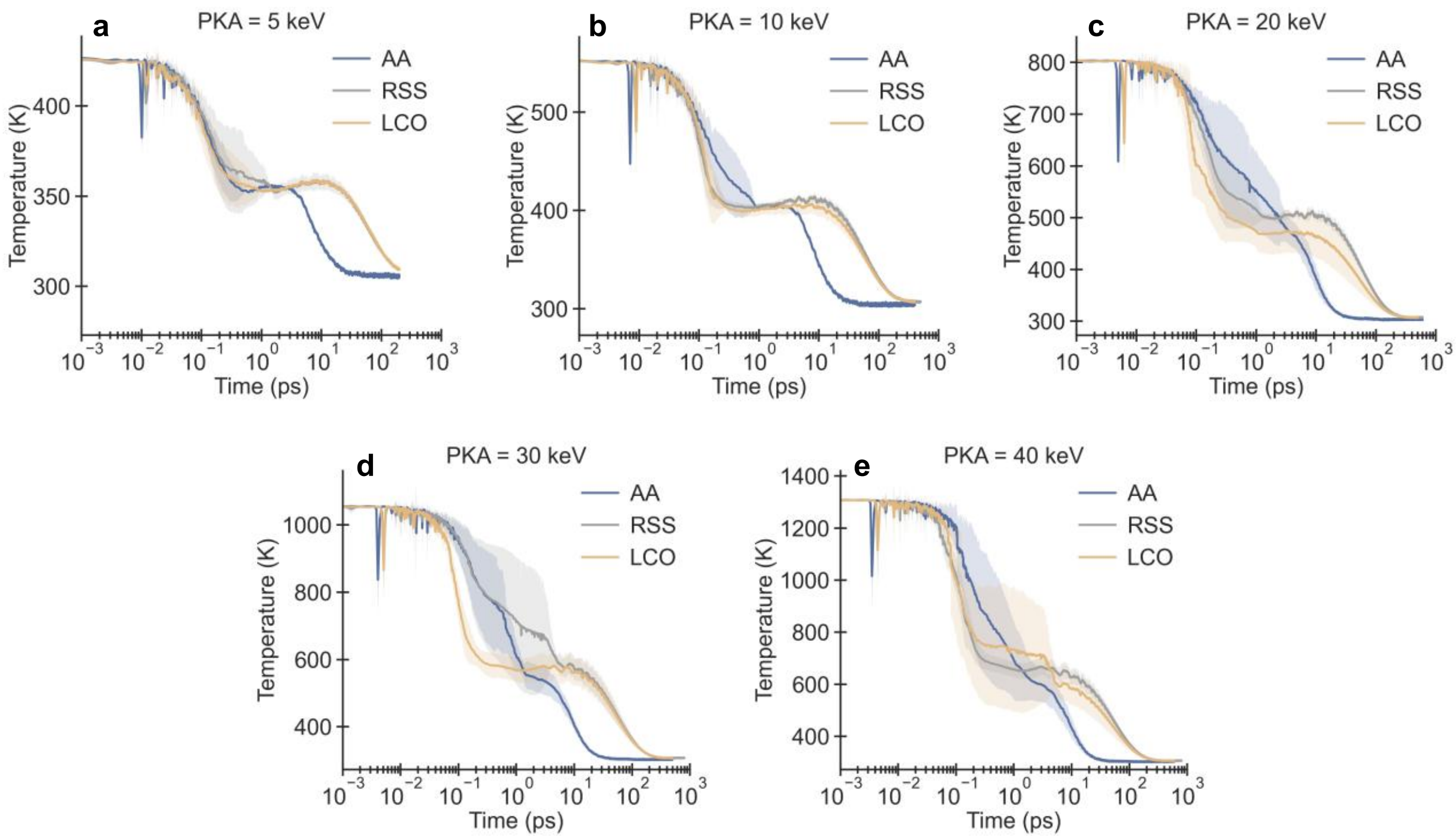

**Fig. 8** The average time evolution of temperature at centered (i.e., NVE) region in AA, RSS and LCO configurations after (a) 5 keV, (b) 10 keV, (c) 20 keV, (d) 30 keV and (e) 40 keV cascades.

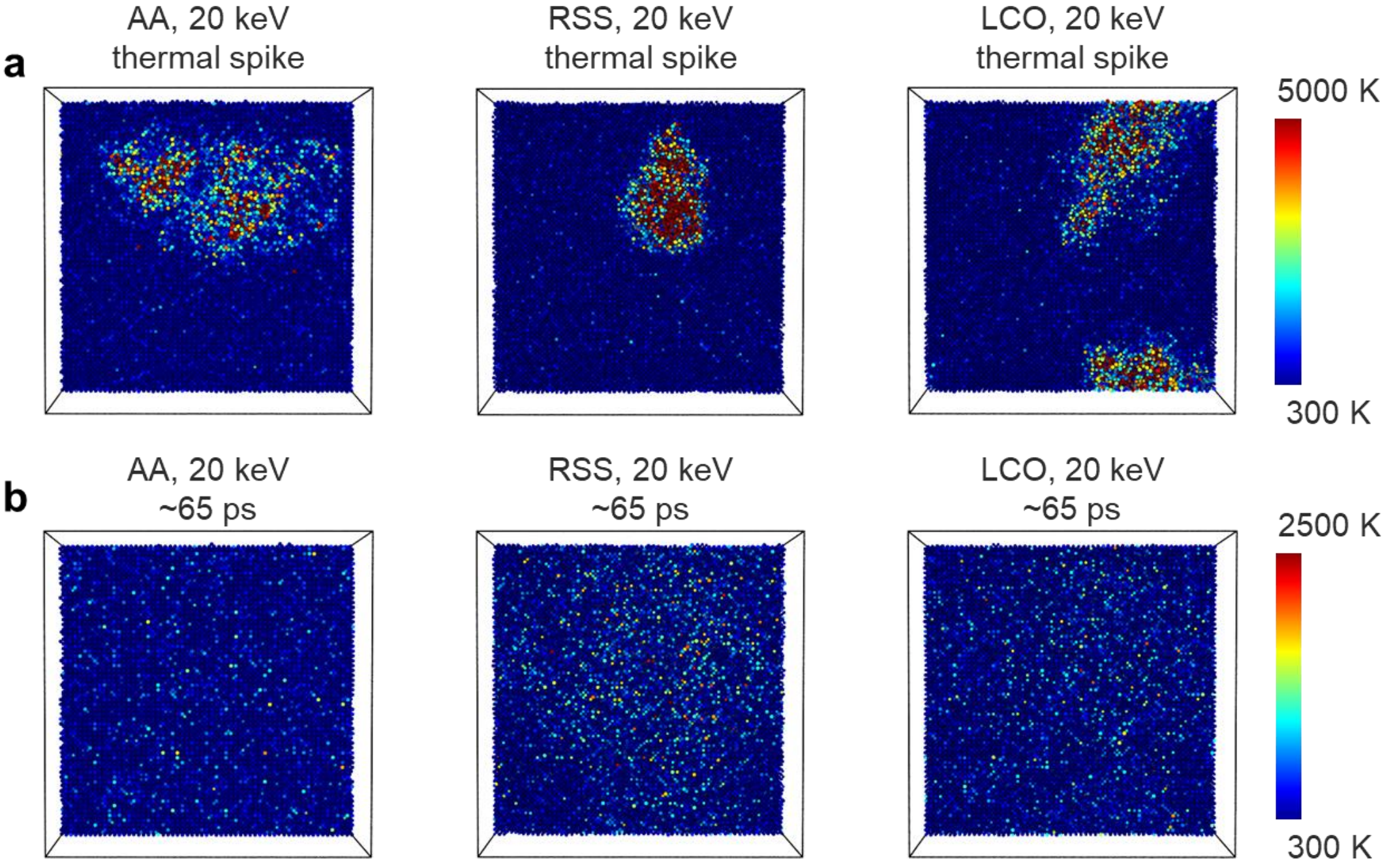

**Fig. 9** Temperature profile (a) at thermal spike and (b) after 65 ps relaxation within AA, RSS and LCO configurations subjected to 20 PKA collision.

Fig. 10 displays the composition of vacancies and interstitials in RSS and LCO models. As shown in Figs. 10a and 10b, the four metals (W, Ta, Cr and V) take up similar atomic fractions of all vacancies, which are mainly within the range of 20%~30% close to their nominal concentration 25 % in the RHEA. The relative fractions show no apparent variation with increasing PKA energy. Such compositions of vacancies can be ascribed to the vacancy formation energies of the four metals, which are within a relatively narrower range (i.e., 2.0~4.0 eV/atom, Table 2) compared with those of interstitials (3.0~10.0 eV/atom, Table 2). On the other hand, interstitials are mainly comprised of Cr atoms. For RSS configurations, the content of interstitial Cr drops from 66 % to 29 % as the PKA energy increases from 5 keV to 40 keV. The higher content of Cr in interstitials under lower PKA energies seems contradict with interstitial formation energy listed in Table 2, showing that V rather than Cr owns the lowest formation energy. In fact, the enrichment of Cr interstitials results from the RSS induced lattice distortion [10]. Table 3 displays the average atomic strain of the four constituent metals in the RSS configurations. Clearly, Cr is subjected to the highest strain level; the severely distorted local atomic arrangement around Cr atoms thereby favors their detachment from original lattice sites under PKA collisions, leading to high concentration of Cr interstitials. As PKA energy increases, the incident energy passed to the RHEA system is so high that all elements at the cascade core are able to overcome lattice cohesive forces and become interstitials, which explains the similar atomic fractions at 40 keV. In contrast to the RSS models, the atomic fraction of interstitial Cr remains high within the entire studied PKA range. This can be attributed to the LCO induced elemental redistribution, which further raises the local distortion and thus strain level of Cr as evidenced in Table 3.

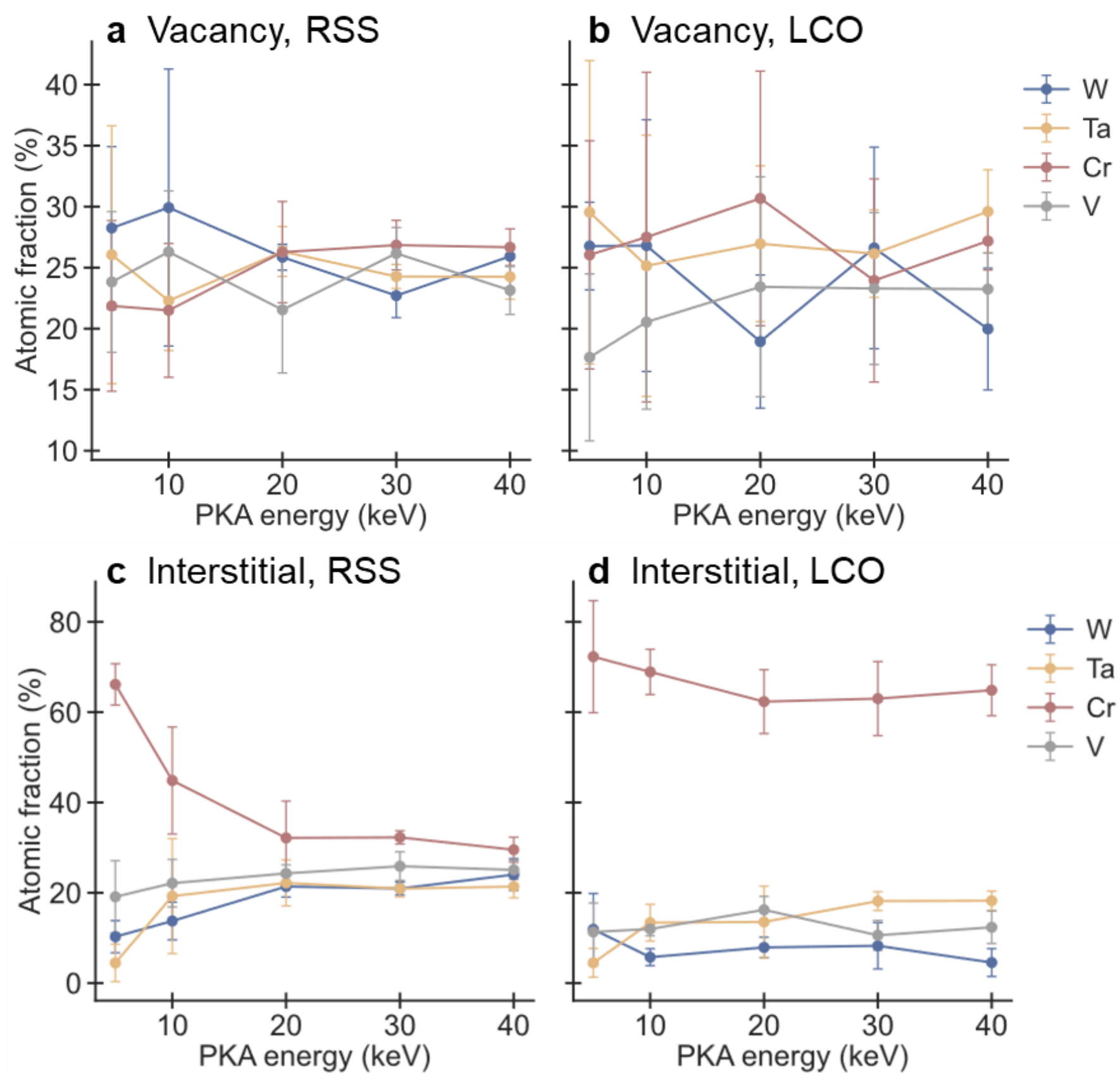

**Fig. 10** Fraction of elements in surviving vacancies and interstitials vs. PKA energy within RSS and LCO configurations.

**Table 2** Point defect formation energies of the four pure metallic materials W, Ta, Cr and V. Values in the parentheses represent DFT results for comparison.

| **Metallic material** | **Point defect formation energy (eV/atom)** | |
|---|---|---|
| | **Vacancy** | **Interstitial** |
| W | 3.90 (3.71) | 9.42 (9.79) |
| Ta | 3.01 (2.86) | 5.94 (5.87) |

| Cr | 2.02 (2.09) | 4.56 (5.18) |
|---|---|---|
| V | 2.74 (2.55) | 3.48 (3.72) |

**Table 3** Average strain level of metals in RSS and LCO configurations. The strain of a specific element is calculated using atomic volume difference in RHEA and in pure metallic phase, $(\sqrt[3]{V_{\text{RHEA}}}-\sqrt[3]{V_{\text{Metallic}}}) / \sqrt[3]{V_{\text{Metallic}}}$.

| **Metallic material** | **Strain (%)** | |
|---|---|---|
| | **RSS** | **LCO** |
| W | -1.459 | -2.231 |
| Ta | -4.808 | -4.469 |
| Cr | 8.570 | 8.857 |
| V | 3.050 | 2.080 |

### 3.3 Mechanical degradation

In this section, the ultimate tensile strength and flow stress of irradiated AA, RSS and LCO configurations under varying PKA energies are studied, as shown in Fig. 11. For configurations unaffected by radiation cascades (i.e., with 0 keV PKA energy), the strength $\sigma_{\text{max}}$ follows the sequence $\sigma_{\text{max}}^{\text{AA}} > \sigma_{\text{max}}^{\text{LCO}} > \sigma_{\text{max}}^{\text{RSS}}$, suggesting that RSS effect weakens alloy strength while LCO effect enhances the resistance against yielding. In fact, this trend of $\sigma_{\text{max}}$ is consistent with previous crystal plasticity theories [15, 18]. It is accepted that the strengths of metals or alloys represent the minimum stress required to activate the nucleation of plastic deformation carriers (e.g., dislocations, deformation twins, secondary phases, etc.) in an initially defect-free crystal. This stress is closely related to internal bond characters that can be qualitatively assessed by the Pugh ratio ($B$ / $G$) and the Cauchy pressure ($C_{12} - C_{44}$), where $B$ and $G$ are the bulk and shear modulus respectively [50, 51]. Lower values of Pugh ratio and Cauchy pressure represent higher covalent bonding characteristics and therefore higher strengths [50, 52]. As listed in Table 4, AA configuration exhibits lowest Pugh ratio and Cauchy pressure and RSS configuration shows highest values (see our previous works [28, 53] for the calculation of $B$ and $G$). Table 4 indicates that $\sigma_{\text{max}}$ of the three configurations should follow the hierarchy of AA > LCO > RSS, which is consistent with MD results in Fig. 11a.

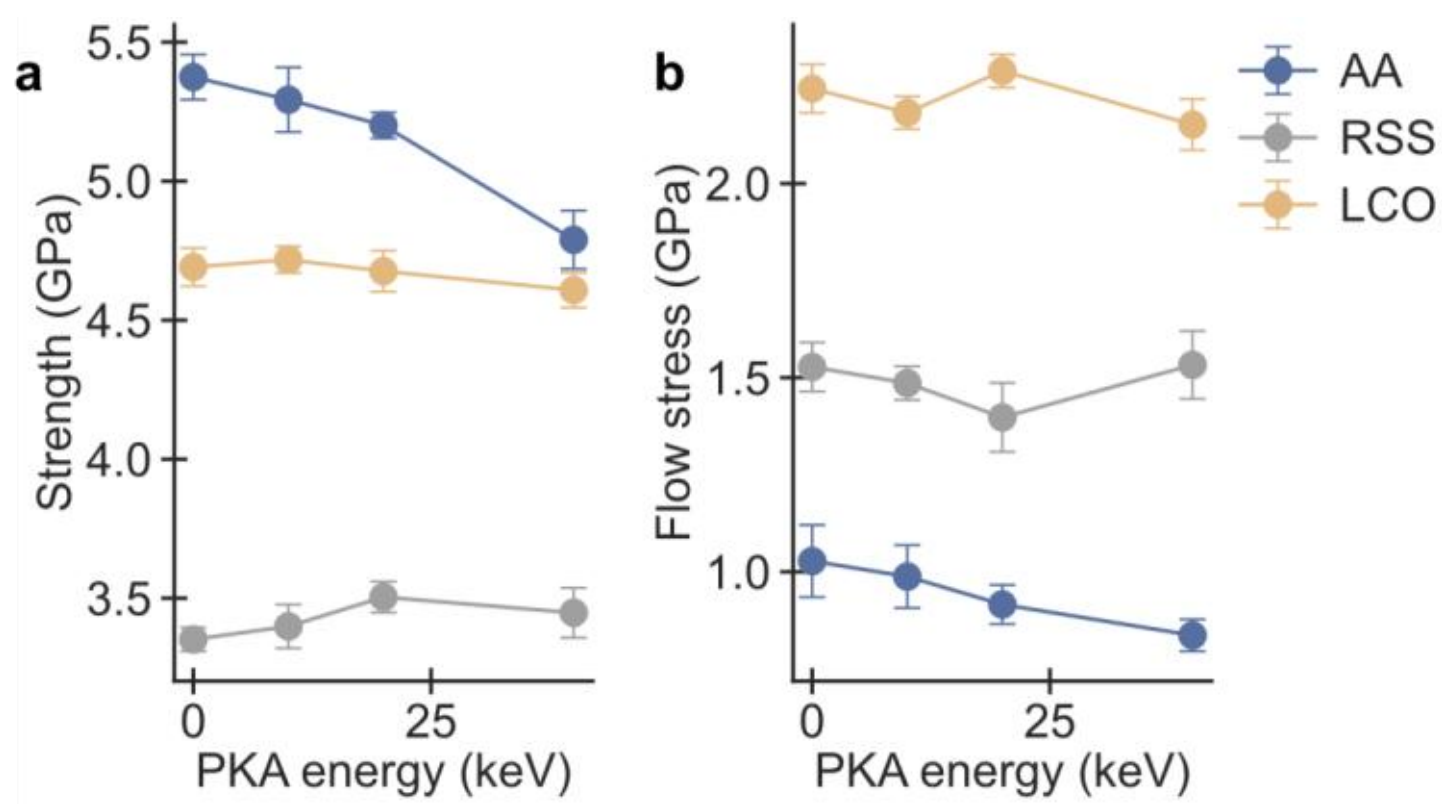

**Fig. 11** (a) Ultimate tensile strength (i.e., maximum values of tensile stress-strain curves) and (b) flow stress (i.e., average stress within 0.15~0.2 strain range) as functions of PKA energy for AA, RSS and LCO RHEA configurations.

**Table 4** The average ratios of bulk modulus $B$ to shear modulus $G$ and the Cauchy pressure $C_{12} - C_{44}$ of AA, RSS and LCO configurations. The results of RSS and LCO configurations represent the mean values of 15 samples.

| Configuration | $B$ / $G$ | $C_{12} - C_{44}$ (GPa) |
|---|---|---|
| AA | 3.064 | 93.6248 |
| RSS | 3.368 | 111.521 |
| LCO | 3.129 | 103.4 |

On the other hand, the flow stress of unirradiated RHEAs follows $\sigma_{\text{flow}}^{\text{LCO}} > \sigma_{\text{flow}}^{\text{RSS}} > \sigma_{\text{flow}}^{\text{AA}}$, which indicates that both RSS and LCO effects elevate the resistance against the movements of nucleated plasticity carriers (i.e., dislocation and twining, as will be discussed subsequently). Comparing the RSS and AA configurations, the interactions among different elements in the former will cause severe lattice distortion and local strain field [10], leading to a roughened potential energy surface with valleys and hills [11, 12]. In this case, the moving plasticity carriers in concentrated alloys always experiences a local energy gradient, which implies an extra dragging force on the moving dislocations or twin boundaries [11]. The above analysis on the RSS-elevated flow stress is supported by the complex shear and volumetric strain fields showing apparent fluctuations due to local composition variation, as presented in Fig. 12. In fact, even for a highly disordered concentrated solution, it is inherently inhomogeneous, because there will always be infinitesimal composition fluctuations much larger than in dilute solutions [11, 13]. Meanwhile, the comparison between the RSS and LCO configurations reveals that LCO effect exerts little impact on local shear strain (Fig. 12), but reduces the frequency of strain fluctuation due to the larger scales of composition domains after elemental pairing (Fig. 3). As mentioned in the previous section, LCO reduces chemical randomness, stabilizes lattice structure and lowers potential energy of the alloy system. As a results, the motion of plasticity carriers, which involves constant breaking and forming local chemical bonds, would require higher external stress to overcome the energetically favored ordering pattern [13, 54]. This explains the highest flow stress level in the LCO configurations.

Considering the radiation effect on mechanical properties, Fig. 11 shows that collision cascades exert little influence on the strength and flow stress of RSS and LCO models, as evidenced by the lack of appreciable variation in these quantities with increasing PKA energy. In contrast, both the strength and flow stress of AA configurations decrease as PKA energy increases. This suggests that, despite the weaker radiation tolerance of the RSS and LCO configurations (Figs. 5 and 6), the RSS and LCO effects can mitigate irradiation-induced mechanical degradation. The suppressed degradation is ascribed to the lattice distortion and internal strain fields in the RSS and LCO models (Fig. 12). As previously reported [1, 55], post-cascade point defects act as preferential nucleation sites for plasticity carriers (such as dislocations, twins or cavities), owing to locally elevated potential energy and disrupted lattice structures. However, because neither large defect clusters nor dislocation loops form after primary damage (Fig. 7), the lattice distortion induced by point defects is confined to a few neighboring shells. As a result, the extent of defect-induced lattice distortion is less significant than the inherent lattice distortion caused by RSS and LCO effects (Fig. 12b). This explains the less pronounced mechanical degradation after primary damage in RSS and LCO configurations relative to the AA configuration.

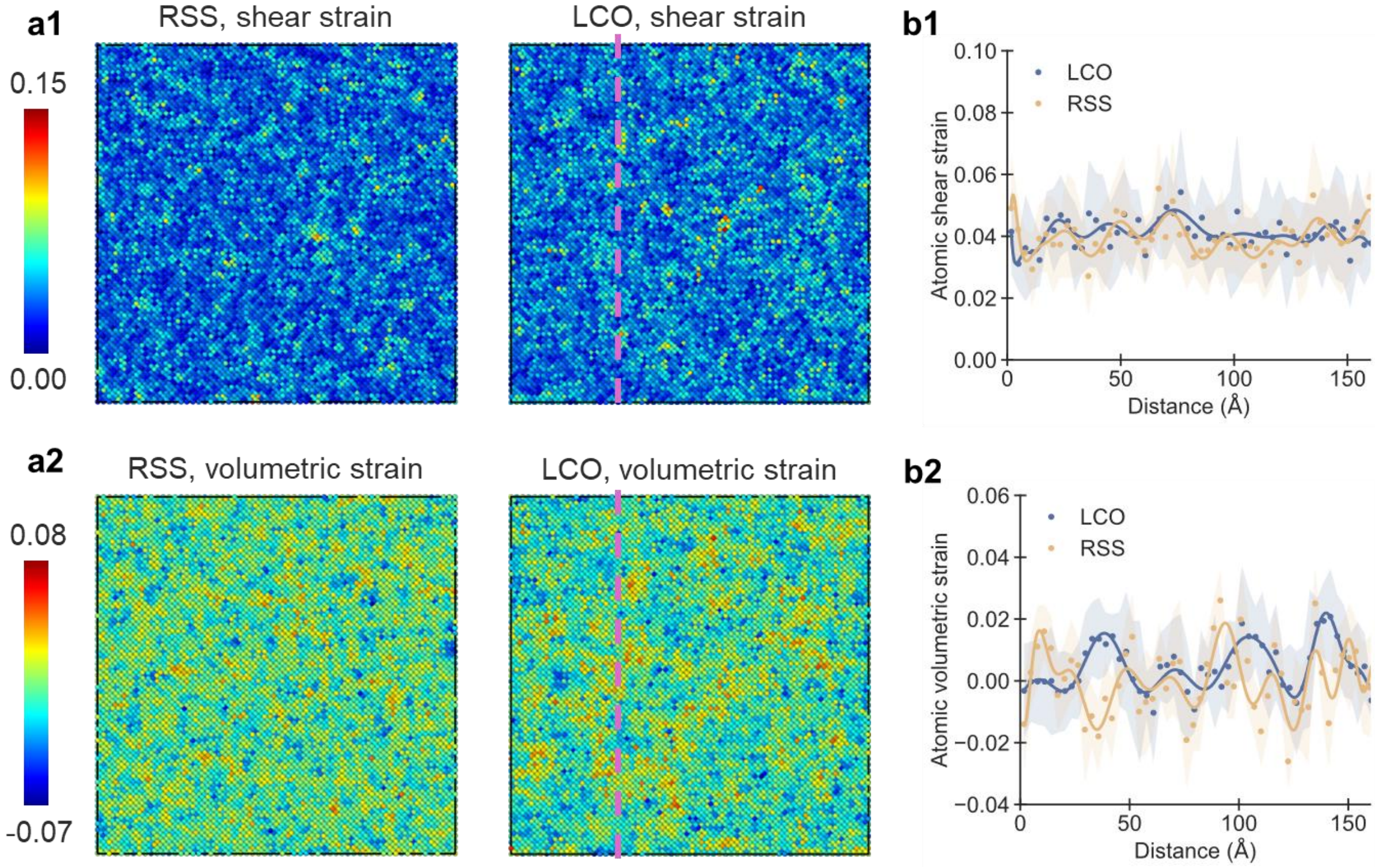

**Fig. 12** (a) The atomic scale strain filed (shear and volumetric) within RSS and LCO RHEA configurations compared with ideal alloy lattice. (b) The corresponding strain profile along dashed lines in (a).

We further investigate the detailed stress-strain responses and nanostructural evolutions of representative AA, RSS and LCO RHEA configurations subjected to 40 keV PKA collisions (Fig. 13). In all three configurations, plastic deformation proceeds mainly through twinning, which is one of the most common deformation mechanisms in BCC metals and alloys and is consistent with previous studies on other BCC RHEAs [4, 26, 56, 57]. During the initial loading stage, all configurations deform elastically (Figs. 13a, 13c and 13e). Once the stress reaches its maximum (i.e., $\sigma_{max}$), nanotwins nucleate and the stored strain energy is abruptly released (Figs. 13b, 13d, 13f). The twining planes are oriented at ~45° relative to the axial direction. In the AA model, twin boundaries (TBs) are mutually parallel; in the RSS and LCO models, TBs intersect each other and restrict their mobility. This is another factor that may contribute to the higher flow stress in these two configurations. In addition, no dislocations are observed in the AA model during the plastic stage, whereas 1/2<111> dislocations appear in the RSS and LCO models. Further analysis on the atomic-scale structural evolution reveals that dislocations nucleate after the TBs sweep through radiation defects (Fig. 13h). Unlike the radiation defects in the AA models (Fig. 13g), defect-TB interactions in the RSS and LCO models exert additional drag forces on the moving TBs, leaving behind dislocation segments. These dislocations are confined to a narrow region between adjacent TBs; their motion is consequently hindered. The restricted dislocations thus lead to accumulated strain energy and a higher flow stress.

It is noteworthy that, according to previous studies, multiple plastic mechanisms can be activated in multicomponent BCC alloys [4, 25, 58, 59], depending on loading conditions, LCO extent, temperature, etc. Here, only two plasticity carriers (i.e., dislocations and twins) are observed during uniaxial tension of the WTaCrV RHEA nanopillars. Based on the developed interatomic potentials, additional plastic mechanisms and the competition among them can be investigated in future work. Moreover, similar to the AA models used in this study, many traditional macro- and meso-scale modelling methods rely on homogenized, averaged materials parameters [60-62] and therefore cannot fully capture the complex chemical inhomogeneities of RHEAs, limiting the accuracy of their predictions. The simulation strategy adopted here explicitly accounts for chemical and structural complexity and reveals the distinct roles of RSS and LCO effects on radiation damage and mechanical behavior. The findings may help identify key physical mechanisms and inform future improvements to macroscale modelling techniques.

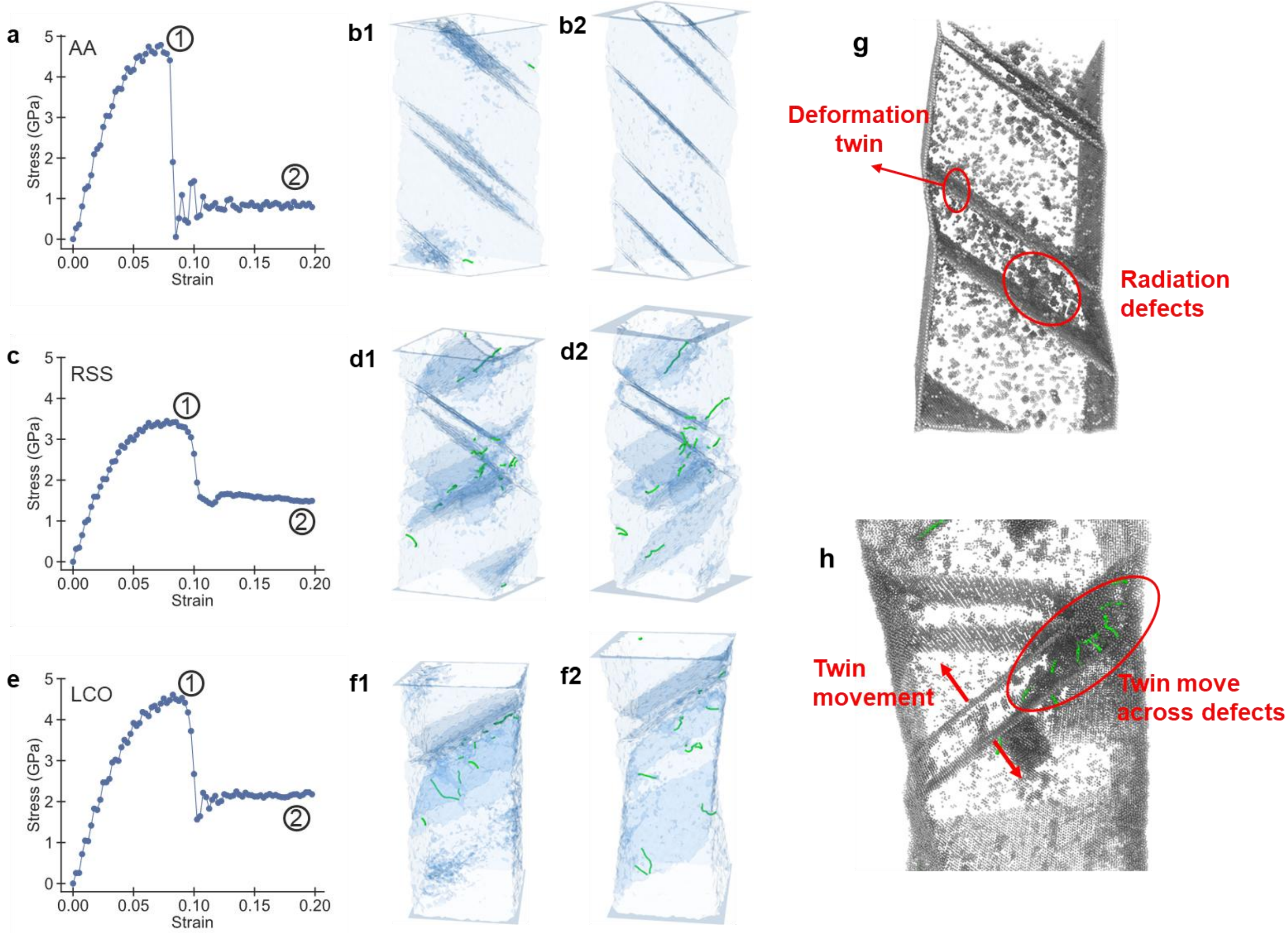

**Fig. 13** (a), (c), (e) The stress-strain curves of AA, RSS and LCO RHEA configurations after 40 keV PKA collision. (b), (d), (f) The 3D-distributions of nanotwins and dislocations. Green lines denote 1/2<111> dislocations. (g) The atomic structure identified using PTM (all BCC atoms are removed), showing interaction between deformation twins and radiation defects in the AA model. (h) The atomic structure showing interaction between twins and radiation defects in the RSS model.

Finally, we investigate whether radiation cascades influence the inherent chemical ordering patterns in the RHEA. As shown in Fig. 14a, the defective RSS model after 40 keV PKA collision is subjected to an additional MC optimization. Results indicate that the Cr content in both vacancies and interstitials continuously increases as MC optimization proceeds (Fig. 14b), eventually approaching the Cr content in radiated LCO configuration (Fig. 10d). This suggests that the presence of radiation defects has little impact on LCO. A subsequent tensile test shows that both the strength and flow stress of this 'RSS+MC' model rise toward those of the LCO model after 40 keV PKA collision. Results in Fig. 14 imply that the radiation-induced mechanical degradation can be mitigated through post-processing treatments such as thermal annealing that are known to modulate the LCO extent. Moreover, under actual service conditions, WTaCrV RHEA is typically exposed to elevated temperatures [7, 44, 63, 64]. Such an environment may promote spontaneous elemental diffusion and LCO, thereby facilitating the recovery of radiation damage and mechanical properties. These findings further highlight the promising applicability and strong potential of WTaCrV in advanced nuclear energy applications.

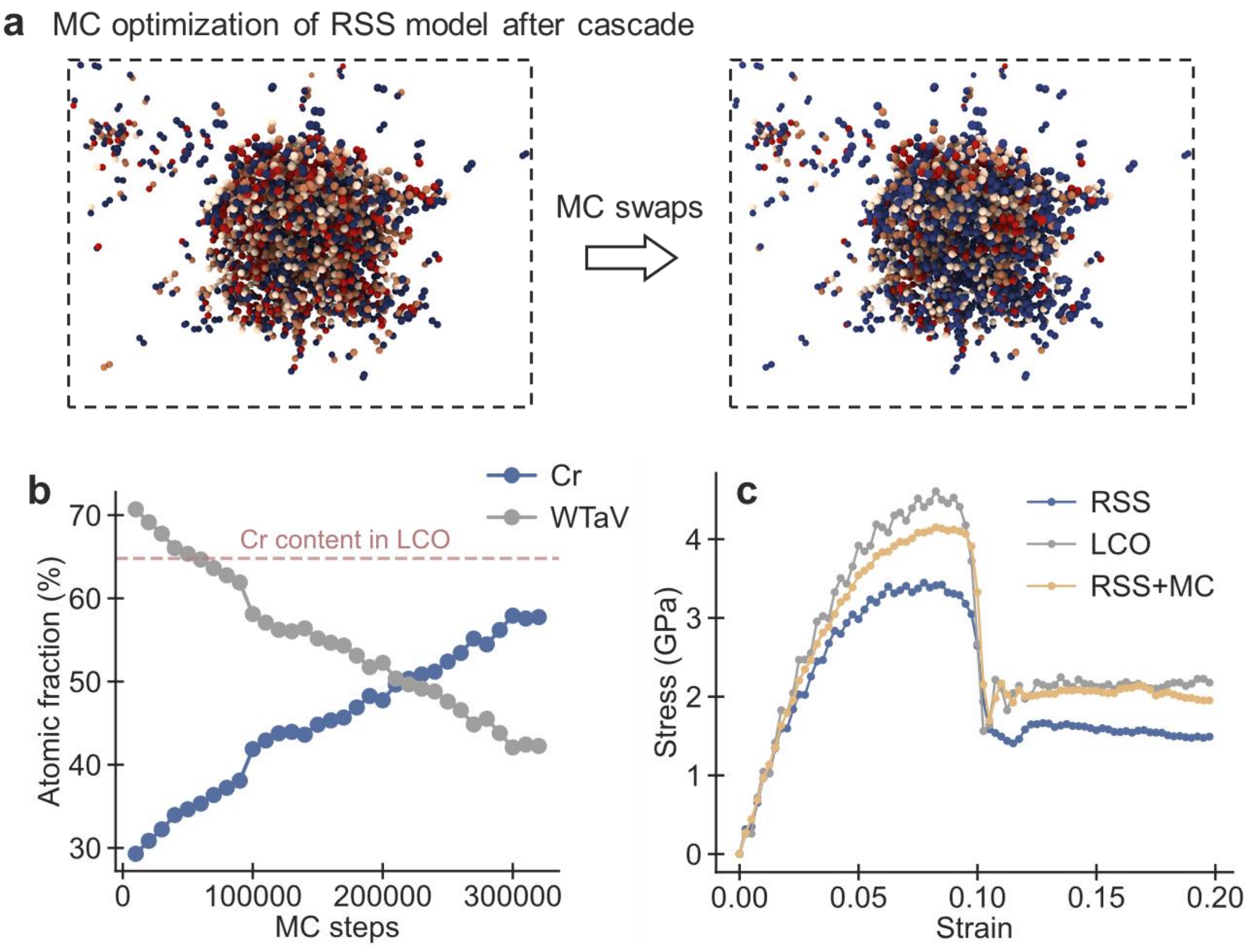

**Fig. 14** (a) The elemental distribution within the cascade region of the RSS model after 40 keV collision, before and after MC optimization. (b) The atomic fraction variation in Cr and the rest three elements during MC optimization. (c) The stress-strain curves of LCO, RSS, and MC optimized RSS models under uniaxial tension.

## 4. Conclusions

This study investigates the effects of RSS and LCO on the primary radiation damage and resulting mechanical degradation of equimolar WTaCrV RHEA via atomistic simulations. The main conclusions are as follows:

1. An atomistic modeling strategy is proposed to isolate and identify the individual roles of solid-solution randomness and short-range chemical ordering on the radiation and mechanical responses of chemically complex alloys.

2. The Alloy and AA interatomic potentials for the WTaCrV RHEA are developed based on DFT reference data. The developed potentials not only accurately reproduce the energetic, structural, and mechanical properties, but also retain excellent structural and thermodynamic stability in dynamic simulations. The accuracy and stability enable its promising applicability in simulating damage evolution and plastic deformation.

3. The primary radiation damage exhibits a strong dependence on chemical environments at nanoscale. The number of radiation defects follows the sequence of $N_{\mathrm{RSS}} > N_{\mathrm{LCO}} > N_{\mathrm{AA}}$. Interactions among different alloy elements in the RSS model cause rugged energy landscape that accelerates defect production. In contrast, the LCO effect enhances lattice cohesion and elevates defect formation energies, thereby effectively suppressing defect accumulation and clustering.

4. In spite of weaker damage resistance of the RSS and LCO configurations under radiation, the RSS and LCO effects are found to suppress mechanical degradation. Uniaxial tensile simulations reveal that collision cascades apparently degrade the mechanical properties of the AA model, but exert a negligible impact on the strength and flow stress of both RSS and LCO models. This mitigation of mechanical degradation is derived from the pronounced inherent lattice distortion and localized strain fields in concentrated alloys, which outweigh defect-induced structural disruptions. Furthermore, the intersecting TBs and the defect-TB interactions in RSS and LCO models restrict plastic flow and promote dislocation storage, leading to higher flow stress.

**Supplementary materials**

Supplementary material associated with this article can be found, in the online version, at https://xxxxxx.

**Data availability**

The data and code that support the findings of this study are available on GitHub at: https://github.com/wuyihan1995/Force_fields_for_refractory_metals.

**Acknowledgements**

The authors acknowledge the support of NSFC (Grant No: 12402212), the Qishan Scholar Project (XRC-25092), and the Computing Center in Xi'an.